\documentclass[letterpaper]{article} % DO NOT CHANGE THIS
\usepackage[draft]{aaai25}  % DO NOT CHANGE THIS
\usepackage{times}  % DO NOT CHANGE THIS
\usepackage{helvet}  % DO NOT CHANGE THIS
\usepackage{courier}  % DO NOT CHANGE THIS
\usepackage[hyphens]{url}  % DO NOT CHANGE THIS
\usepackage{graphicx} % DO NOT CHANGE THIS
\usepackage{natbib}  % DO NOT CHANGE THIS AND DO NOT ADD ANY OPTIONS TO IT
\usepackage{caption} % DO NOT CHANGE THIS AND DO NOT ADD ANY OPTIONS TO IT
\usepackage{algorithm}
\usepackage{algorithmic}
\usepackage{booktabs}
\usepackage{multirow}
\usepackage{amsmath}
\usepackage{amsfonts}
\usepackage{amsthm}
\usepackage{amssymb}

\usepackage{newfloat}
\usepackage{listings}
\DeclareCaptionStyle{ruled}{labelfont=normalfont,labelsep=colon,strut=off} % DO NOT CHANGE THIS
\floatstyle{ruled}
\newfloat{listing}{tb}{lst}{}
\floatname{listing}{Listing}
\title{Lightweight yet Fine-grained: A Graph Capsule Convolutional Network with Subspace Alignment for Shared-account Sequential Recommendation}
\author{
    Jinyu Zhang\textsuperscript{\rm 1},
    Zhongying Zhao\textsuperscript{\rm 1}\thanks{Corresponding Author.},
    Chao Li\textsuperscript{\rm 1}, Yanwei Yu\textsuperscript{\rm 2}
}
\affiliations{
    \textsuperscript{\rm 1}College of Computer Science and Engineering, Shandong University of Science and Technology, Qingdao 266590, China\\
    \textsuperscript{\rm 2}College of Computer Science and Technology, Ocean University of China, Qingdao 266400, China\\
    jinyuz1996@outlook.com, zzysuin@163.com or  zyzhao@sdust.edu.cn, lichao@sdust.edu.cn, yuyanwei@ouc.edu.cn
}

\usepackage{bibentry}
\begin{document}

\maketitle

\begin{abstract}
Shared-account Sequential Recommendation (SSR) aims to provide personalized recommendations for accounts shared by multiple users with varying sequential preferences. Previous studies on SSR struggle to capture the fine-grained associations between interactions and different latent users within the shared account's hybrid sequences. Moreover, most existing SSR methods (e.g., RNN-based or GCN-based methods) have quadratic computational complexities, hindering the deployment of SSRs on resource-constrained devices. To this end, we propose a \textbf{Light}weight \textbf{G}raph \textbf{C}apsule \textbf{C}onvolutional \textbf{N}etwork with subspace alignment for shared-account sequential recommendation, named LightGC$^2$N. Specifically, we devise a lightweight graph capsule convolutional network. It facilitates the fine-grained matching between interactions and latent users by attentively propagating messages on the capsule graphs. Besides, we present an efficient subspace alignment method. This method refines the sequence representations and then aligns them with the finely clustered preferences of latent users. The experimental results on four real-world datasets indicate that LightGC$^2$N outperforms nine state-of-the-art methods in accuracy and efficiency. 
% The code and dataset of this work are available at \url{https://github.com/ZZY-GraphMiningLab/LightGC2N}.
\end{abstract}

\begin{links}
    \link{Code}{https://github.com/ZZY-GraphMiningLab/LightGC2N}
    % \link{Datasets}{https://aaai.org/example/datasets}
    % \link{Extended version}{https://aaai.org/example/extended-version}
\end{links}

\section{Introduction}
\begin{figure}[ht]
\centering
\includegraphics[width=1.0\columnwidth]{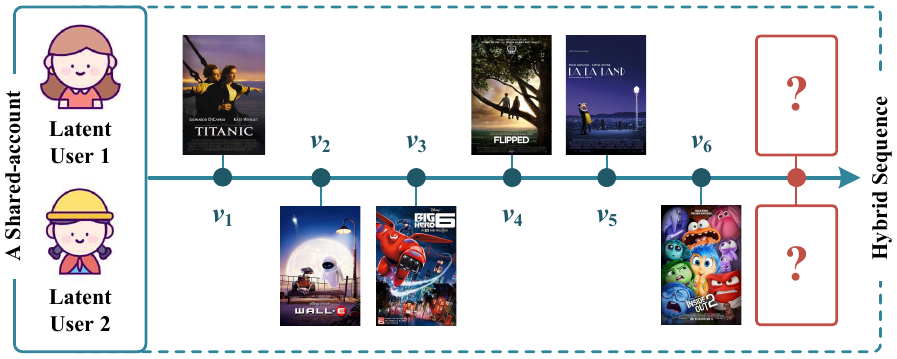} % Reduce the figure size so that it is slightly narrower than the column. Don't use precise values for figure width.This setup will avoid overfull boxes.
\caption{An example to illustrate the shared-account sequential recommendation scenario. $\{ v_1, v_2, \dots, v_6\}$ are the historical behaviors in the hybrid sequence.}
\label{fig1:motivation}
\end{figure}

% Para.1
Sequential Recommender systems (SRs) strive to provide users with personalized content \cite{Ma_Sequential_24_AAAI}, products \cite{Yue_Graph_23_TBD}, or services \cite{Li_SR_24_AAAI} based on their sequential preferences. Most SRs are under an ideal assumption that each account is merely associated with one single user \cite{Verstrepen_Shared_15_RecSys}. In real-world scenarios, many users prefer to share their accounts with family members or close friends \cite{Jiang_Shared_18_SIGIR}. 
As illustrated in Figure \ref{fig1:motivation}, multiple family members (i.e., latent users) utilize a shared video account. 
% Despite their distinct preferences, the watching logs of these users are intermingled within an account-level hybrid sequence.
Clearly, their viewing histories reveal distinct preferences, yet they’re blended into an account-level hybrid sequence.
Distinguishing the diverse preferences of latent users while providing account-level sequential recommendations emerges as an appealing yet challenging task, i.e., the Shared-account Sequential Recommendation (SSR).
% Note that, account sharing is a positive, proactive behavior between close friends or family members \cite{Guo_Shared_23_TKDE}. Instead of linking two or more accounts that may belong to the same person, SSR aims to provide more accurate and reliable recommendations for known shared accounts without collecting additional information \cite{Guo_Shared_23_TKDE}. Consequently, SSR does not pose a risk of exposing user privacy.

% Para.2
In the early explorations of the SSR, researchers prefer utilizing Recurrent Neural Networks (RNNs) to distinguish the user preferences from shared accounts \cite{Wen_Shared_21_DASFAA}. $\pi$-Net \cite{Ma_Shared_19_SIGIR} and PSJNet \cite{Sun_Shared_23_TKDE} are two RNN-based SSR methods that leverage Gating Recurrent Unit (GRU) to cluster the preferences of latent users and then learn the account-level sequence representations. Since an account encompasses multiple latent users, the shared-account sequences tend to be longer than typical sequences. However, RNN-based methods have catastrophic forgetting problems in processing lengthy sequences, leading to gradient vanishing problems during model training \cite{Guo_Shared_24_TNNLs}.
Subsequently, DA-GCN \cite{Guo_Shared_21_IJCAI} and TiDA-GCN \cite{Guo_Shared_24_TNNLs} are proposed. These graph-based SSR methods model the preferences of each latent user through a multi-head self-attention mechanism and learn account-level sequence embeddings by propagating messages on the sequential graph.
Although these methods have achieved remarkable performance in SSR, they still face the below two challenges:

% Para.3
\textbf{(1) Coarse-grained user representations.} Since an account is shared by multiple users in SSR, the interactions are generated by diverse latent users. Most graph-based SSR methods directly learn the preferences of latent users by modeling the account-level sequences on the sequential graphs \cite{Wen_Shared_21_DASFAA}. They try to perform clustering \cite{Guo_Shared_21_IJCAI} or apply multi-head self-attention \cite{Guo_Shared_24_TNNLs} to extract the multiple interests of each account, thereby simulating the diverse preferences of latent users.
However, these methods fail to distinguish the ownership of each interaction within the hybrid sequences, making it challenging for SSRs to capture the fine-grained preferences of latent users. 

% Para.4
\textbf{(2) High computational complexity.} The graph-based shared-account sequential recommenders usually integrate complicated structures (e.g., self-attention \cite{Guo_Shared_24_TNNLs} or clustering methods \cite{Sun_Shared_23_TKDE, Jiang_Shared_18_SIGIR}) to differentiate the preferences of multiple latent users from hybrid sequences. However, both the self-attention mechanism \cite{Katharopoulos_LinearAtt_20_ICML} and the clustering method based on self-representation matrices \cite{Cai_subspace_22_CVPR} have quadratic computational complexity. Such a high computational demand seriously impacts the user experience and poses a challenge for deploying graph-based SSRs on resource-constrained mobile devices.

% Para.5
To tackle the above problems, we propose a \textbf{Light}weight \textbf{G}raph \textbf{C}apsule \textbf{C}onvolutional \textbf{N}etwork (LightGC$^2$N). Specifically, we present a lightweight Graph Capsule Convolutional Network (GC$^2$N) to identify the ownership of each interaction for latent users. In this component, we construct capsule graphs to identify the ownership of interactions for different latent users. By attentively propagating messages on the graphs, GC$^2$N performs a fine-grained distinction of the preferences for different latent users. Furthermore, we design an account-level dynamic routing mechanism. It merges the preferences of latent users, yielding the account-level capsule representations. Besides, we devise an efficient Subspace Alignment (SA) method that utilizes low-rank subspace bases to refine the sequence embeddings. Finally, SA adopts a contrastive learning strategy to align the preferences of latent users between refined and original sequences.
% Finally 

% para. 6 Main contributions
The main contributions of this work includes:
\begin{itemize}
    \item We propose a subspace alignment-enhanced graph capsule convolutional network for the shared-account sequential recommendation, namely LightGC$^2$N.
    \item We design a lightweight graph capsule convolutional network that finely distinguishes the preferences of each latent user within an account.
    \item We devise an efficient subspace alignment method that refines the sequence embeddings and aligns them with the preferences of latent users.
    \item Experimental results on four datasets demonstrate that LightGC$^2$N outperforms other state-of-the-art SSR methods in terms of performance and model efficiency.
\end{itemize}

\section{Related Work}
\subsection{Sequential Recommendation}
Sequential recommender systems (SRs) predict users’ next interactions based on their sequential preferences \cite{Chen_SR_24_AAAI}. Early researches utilized Markov chains to address the sparsity issues in SR tasks \cite{He_SR_16_ICDM, Cai_SR_17_IJCAI}, but they fail to capture the dynamics of user preferences. Subsequently, researchers begin to explore deep neural networks for SRs, including RNN-based methods \cite{Quadrana_SR_17_RecSys}, GNN-based methods \cite{Fan_SR_21_CIKM}, attention-based methods \cite{Kang_SR_18_ICDM, Shin_SR_24_AAAI, He_SR_18_TKDE}, and contrastive learning-based methods \cite{Xie_SR_22_ICDE}. These deep learning-based SR methods excel in capturing dynamic sequential patterns of users. Nevertheless, they typically assume that each account is associated with a single user \cite{Guo_Shared_23_TKDE}, thus failing to provide accurate recommendations for shared accounts.

\subsection{Shared-account Sequential Recommendation}
Shared-account Sequential Recommender systems (SSRs) aim to identify the diverse preferences of latent users while providing personalized recommendations for shared accounts \cite{Verstrepen_Shared_15_RecSys}. Early SSR research utilized RNN-based methods \cite{Ma_Shared_19_SIGIR, Sun_Shared_23_TKDE} to identify latent users, employing GRU units to filter out information from each account. However, these methods suffer from gradient vanishing issues with long sequences. Subsequently, the graph-based SSR methods with attention networks are proposed \cite{Guo_Shared_21_IJCAI, Guo_Shared_24_TNNLs}, which propagates user-specific messages to identify latent users.
These graph-based methods have achieved remarkable performance on SSR. However, they didn't consider the fine-grained associations between interactions in sequences and latent users. Besides, their high computational complexity hinders the deployment of SSRs on resource-constrained edge devices (e.g., smartphones or tablets). 

\begin{figure*}[ht]
\centering
\includegraphics[width=2\columnwidth]{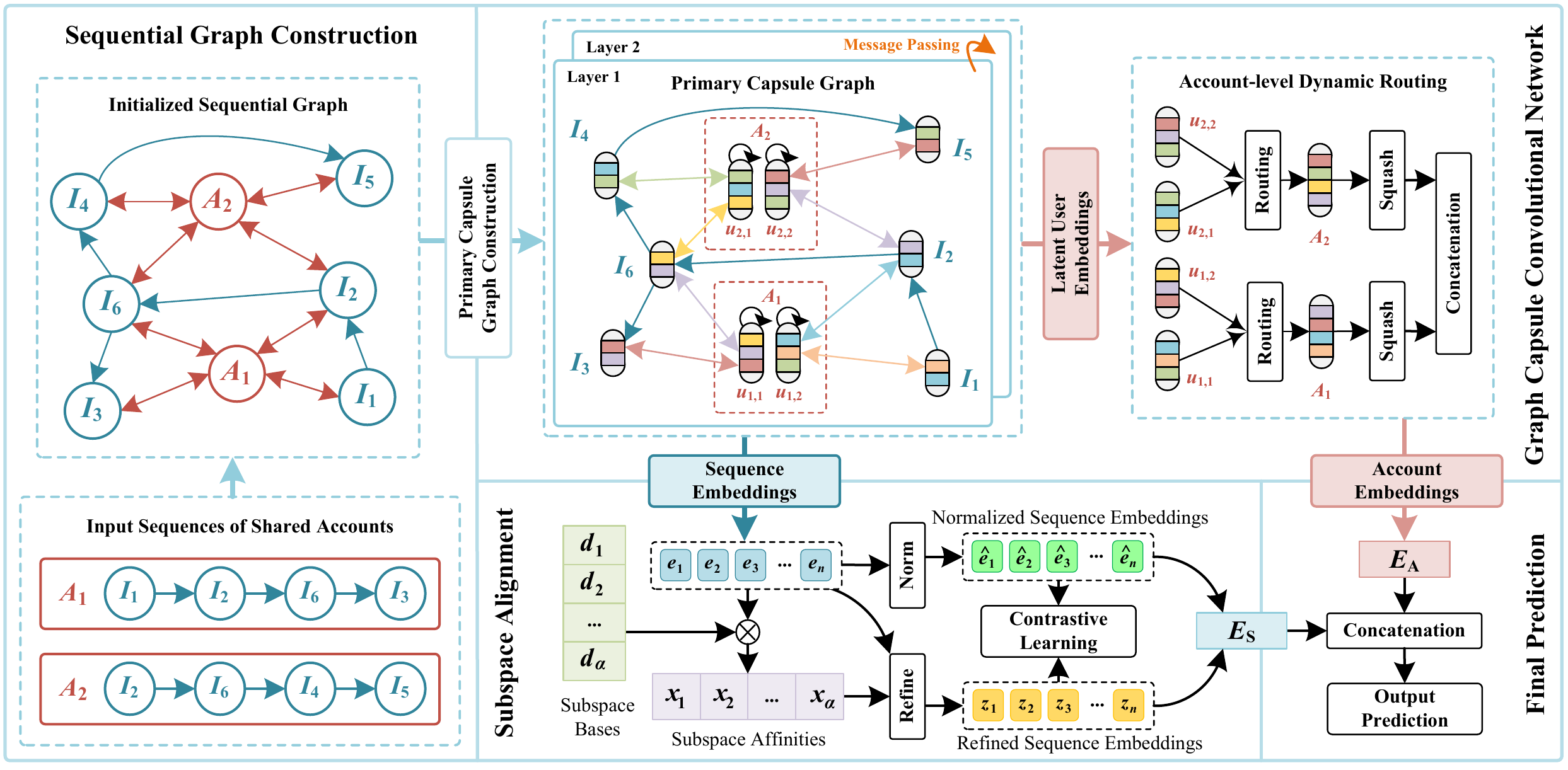} % Reduce the figure size so that it is slightly narrower than the column. Don't use precise values for figure width.This setup will avoid overfull boxes.
\caption{Framework of LightGC$^2$N, where $A_1$ and $A_2$ represent two shared accounts, and
$\{ I_1, I_2, \dots, I_6\}$ denote the historical interactions that compose the hybrid sequences for these accounts.}
\label{fig2:model}
\end{figure*}

\section{Methodology}
\subsection{Preliminaries}
\subsubsection{Notations.} Suppose that $\mathcal{I} = \{I_1, I_2, \dots, I_t, \dots, I_m\}$ is the set of items, where $I_t$ denotes the $t$-th item. The set of shared accounts is denoted as $\mathcal{A} = \{A_1, A_2, \dots, A_k, \dots, A_n\}$, where $A_k$ represents the $k$-th shared account. Moreover, the set of sequences is denoted as $\mathcal{S} = \{S_1, S_2, \dots, S_k, \dots, S_n\}$, where $S_k$ denotes the hybrid sequence of the shared account $A_k$. Suppose that each account contains $\alpha$ latent users, e.g., $A_k = \{u_{k,1}, \dots, u_{k,h}, \dots, u_{k,\alpha}\}$, where $u_{k,h}$ denotes the $h$-th latent user in the account $A_k$. 
% Suppose $I_j \in \mathcal{N}^{u_{k,h}}_{I}$, then $I_j$ is an interacted item of $u_{k,h}$, where $\mathcal{N}^{u_{k,h}}_{I}$ denotes the set of all interacted items of $u_{k,h}$. The $\mathcal{N}^{I_j}_{u}$ is the set of latent users who have interacted with item $I_j$. $\mathcal{N}^{A_k}_{I}$ is the set of all the interacted items of $A_k$. Additionally, the set of neighbor items for $I_j$ is denoted as $\mathcal{N}^{I_j}_{I}$.

\subsubsection{Problem Definition.}
Given $S_k$ and $A_k$, the task of SSR is to recommend the next item $I_{t+1}$ that $A_k$ is most likely to consume, based on the account's hybrid sequence $S_k$. The probabilities of all recommendation candidates are represented as:
\begin{align}
    P(I_{t+1} | S_k, A_k) \sim f(S_k, A_k),
\end{align}
where $P(I_{t+1} | S_k, A_k)$ denotes the probability of recommending $I_{t+1}$ to $A_k$ given its historical hybrid sequence $S_k$, and $f(S_k, A_k)$ is the function designed to estimate the probability.

\subsection{Framework of LightGC$^2$N}
As shown in Figure \ref{fig2:model}, the LightGC$^2$N consists of four key components: 1) sequential graph construction, 2) graph capsule convolutional network, 3) subspace alignment, and 4) final prediction. The details are given below.

\subsection{Sequential Graph Construction} 
The sequential graphs are constructed as the input for the graph capsule convolutional network. During the graph construction, two types of associations are considered: account-item interactive relations and sequential dependencies between items. We define the sequential graphs as $\mathcal{G} = \{\mathcal{V}, \mathcal{E}\}$, where $\mathcal{V}$ is the set of nodes, and $\mathcal{E}$ is the set of edges. Each edge in $\mathcal{E}$ denotes a relation between nodes.
Specifically, the adjacency matrix $\mathcal{M} \in \mathbb{R}^{(m+n) \times (m+n)}$ of the sequential graph $\mathcal{G}$ is denoted by Eqn. (\ref{eqn:seq_graph}).
\begin{equation}
    \mathcal{M} = \begin{bmatrix}
        \mathbf{M}_{S} & \mathbf{M}_{I}\\
        \mathbf{M}_{I}^\top & \mathbf{0}\\
    \end{bmatrix}, \label{eqn:seq_graph}
\end{equation}
where $\mathbf{M}_{S} \in \mathbb{R}^{m \times m}$ denotes the adjacency matrix carrying the sequential relationships between items, and each entry $M_{ij} = 1$ if item $I_j$ is the prior of item $I_i$ in the input sequences; $M_{ij} = 0$ otherwise. $\mathbf{M}_{I} \in \mathbb{R}^{m \times n}$ represents the adjacency matrix containing the interactive relationships between accounts and items, each entry $M_{kl} = 1$ if account $A_k$ has interaction with item $I_l$; $M_{kl} = 0$ otherwise.

\subsection{Graph Capsule Convolutional Network (GC$^2$N)}
We design the GC$^2$N is to capture the fine-grained preferences of each latent user within a shared account. The inputs to GC$^2$N are the associations among nodes from the initialized sequential graphs and the embeddings at the 0-th layer, i.e., $\mathbf{E}_I^{(0)} \in \mathbb{R}^{m \times d_1}$ for all items and $\mathbf{E}_A^{(0)} \in \mathbb{R}^{n \times d_1}$ for all accounts.

\subsubsection{Primary Capsule Graph Construction.} 
% We first construct the primary capsule graphs. Capsules are high-dimensional embeddings that is able to carry multi-aspect information of a single object \cite{Wu_Capsule_Graph_23_TNNLs}. 

In this component, node embeddings are projected into high-dimensional capsule spaces to construct the primary capsule graphs, thereby exploring the fine-grained differences in preferences between latent users.
Specifically, this component utilizes linear attention mechanism \cite{Katharopoulos_LinearAtt_20_ICML} to project item embeddings $\mathbf{E}_I^{(0)}$ into item capsules $\mathbf{C}_I^{(0)} \in \mathbb{R}^{m \times d_2}$, where the $\mathbf{E}_I^{(0)}$ is treated as \textbf{Q}uery, \textbf{K}ey and \textbf{V}alue during the calculation. In contrast to the self-attention mechanism, the linear attention mechanism first calculates the outer product between the \textbf{K}ey matrix and the \textbf{V}alue matrix to obtain the attention map, and then calculates the correlations between the map and the \textbf{Q}uery matrix. Since the linear attention changes the calculation orders, it achieves lower computational complexity (i.e., $\mathcal{O}(N \times d^2)$) while capturing global associations among items. The calculation of the linear attention mechanism is formulated as Eqn. (\ref{eqn:LA}).
\begin{equation}
      \mathbf{C}_I^{(0)} =  \mathbf{Q} \left( \text{softmax} \left(\frac{{\mathbf{K}}^\top \mathbf{V}}{\sqrt{d_1}} \right) \right) \mathbf{W}_l + \mathbf{b}_l, \label{eqn:LA}
\end{equation}
where $\mathbf{W}_l \in \mathbb{R}^{d_1 \times d_2}$ is a dimension transformation matrix, and $\mathbf{b}_l \in \mathbb{R}^{1 \times d_2}$ is the bias term.

Since there are no sequential relationships between accounts, it does not need to use an attention-based method for account representations $\mathbf{E}_A^{(0)}$. Instead, GC$^2$N leverages point-wise Conv1D to project them into capsule-form $\mathbf{C}_A^{(0)} \in \mathbb{R}^{n \times \alpha \times d_2}$. In the point-wise Conv1D, both the kernel size and stride are set to 1 for linear dimensional transformation \cite{Wu_Capsule_Graph_23_TNNLs}. The calculation is denoted as Eqn. (\ref{eqn:conv1d}).
\begin{equation}
    \mathbf{C}_A^{(0)} = \mathbf{E}_A^{(0)} \ast \mathbf{W}_c + \mathbf{b}_c, \label{eqn:conv1d}
\end{equation}
where $\ast$ represents the convolutional operation, $\mathbf{W}_c \in \mathbb{R}^{d_1 \times \alpha \times d_2}$ denotes the kernel of Conv1D, $\mathbf{b}_c \in \mathbb{R}^{1 \times \alpha \times d_2}$ represents the bias, and $\alpha$ is a hyper-parameter that controls the number of latent users within shared accounts.

Subsequently, each account capsule $\mathbf{C}_{A_k}^{(0)} \in \mathbb{R}^{\alpha \times d_2}$ is split into $\alpha$ latent user capsules $\mathbf{C}_{u_k}^{(0)} \in \mathbb{R}^{d_2}$. Then, the model is able to match the interactions to the specific latent users within an account by attentively calculating the correlations between their capsules, which facilitates the fine-grained distinction of the preferences for latent users.

\subsubsection{Fine-grained Message Propagation on capsule graphs.}
Taking the shared-account $A_k$ as an example. The capsule embedding of the $h$-th latent user in $A_k$ is denoted as $\mathbf{C}_{u_{k, h}}$.
To realize fine-grained message passing on the primary capsule graphs, GC$^2$N calculates the correlations $a_{I_j}$ between items $\mathbf{C}_{I_{j}}$ and user $\mathbf{C}_{u_{k,h}}$ as:
\begin{equation}
    a_{I_j} = \frac{\exp\left( \mathbf{C}_{u_{k,h}} \cdot \mathbf{C}_{I_{j}} \right)}{\sum_{I_j \in \mathcal{N}^{A_k}_{I}} \sqrt{\exp\left( \mathbf{C}_{u_{k,h}} \cdot \mathbf{C}_{I_{j}} \right)}}, \label{eqn:attention_pro}
\end{equation}
where $\mathcal{N}^{A_k}_{I}$ is the set of all the interacted items of $A_k$.

Such an attentive calculation facilitates the fine-grained matching between interactions and latent users.
Then, the messages propagated to $u_{k,h}$ at the $l$-th layer are denoted as:
\begin{equation}
    \mathbf{m}_{u_{k,h}\leftarrow I_{j}}^{(l)}=\mathbf{W}_{1}^{(l)}\mathbf{C}_{I_{j}}^{(l-1)}+\mathbf{W}_{2}^{(l)} (a_{I_j}^{(l-1)} \mathbf{C}_{I_j}^{(l-1)}), \label{eqn:message_iu}
\end{equation}
where $\mathbf{m}_{u_{k,h}\leftarrow I_{j}}^{(l)}$ denotes the passed message, $\mathbf{W}_{1}^{(l)}$ denotes the learnable weights that controls how much information should be passed from neighboring item $I_j$, and $\mathbf{W}_{2}^{(l)}$ is another learnable weighting matrix that controls the participation of correlations.

We also add self-connections to retain the independent characteristics of $\mathbf{C}_{u_{k,h}}^{(l)}$, which is formulated as:
\begin{equation}
    \mathbf{m}^{(l)}_{u_{k,h}\leftarrow u_{k,h}}=\mathbf{W}_{3}^{(l)}\mathbf{C}_{u_{k,h}}^{(l-1)}, \label{eqn:meesage_uu}
\end{equation}
where $\mathbf{m}^{(l)}_{u_{k,h}\leftarrow u_{k,h}}$ is the retained information of the user capsule from $(l-1)$-th layer, $\mathbf{W}_3^{(l)}$ is the learnable parameter that controls how much information of $\mathbf{C}_{u_{k,h}}^{(l-1)}$ should be retained.

Hence, the capsule representation $\hat{\mathbf{C}}_{u_{k,h}}^{(l)}$ of latent user $u_{k,h}$ is updated by Eqn.~(\ref{eqn:message_aggregate_u}).
\begin{equation}
   \hat{\mathbf{C}}_{u_{k,h}}^{(l)}=\sum_{I_{j}\in\mathcal{N}^{u_{k,h}}_{I}}\mathbf{m}^{(l)}_{u_{k,h}\leftarrow I_{j}} + \mathbf{m}^{(l)}_{u_{k,h}\leftarrow u_{k,h}},\label{eqn:message_aggregate_u}
\end{equation}
where $\mathcal{N}^{u_{k,h}}_{I}$ denotes the set of all items interacted by $u_{k,h}$.

Similarly, the message propagated to item capsule $\mathbf{C}_{I_j}$ at $l$-th layer is represented by Eqn. (\ref{eqn:meesage_i}):
\begin{align}
    \mathbf{m}^{(l)}_{I_{j}\leftarrow u_{k,g}}=\mathbf{W}^{(l)}_{4}\mathbf{C}^{(l-1)}_{u_{k,g}}; \quad
    \mathbf{m}^{(l)}_{I_{j}\leftarrow I_{j-1}}=\mathbf{W}^{(l)}_{5}\mathbf{C}^{(l-1)}_{I_{j-1}},\label{eqn:meesage_i}
\end{align}
where $u_{k,g}$ represents a latent user who has interacted with $I_j$, $\mathbf{m}^{(l)}_{I_{j}\leftarrow u_{k,g}}$ is the message passed from $u_{k,g}$ to $I_j$, and $I_{j-1}$ is the neighboring item of $I_j$, $\mathbf{m}^{(l)}_{I_{j}\leftarrow I_{j-1}}$ denotes the message passed from $I_{j-1}$ to $I_{j}$, $\mathbf{W}^{(l)}_4$ and $\mathbf{W}^{(l)}_5$ are learnable weights.

Then, the capsule representation $\hat{\mathbf{C}}^{(l)}_{I_j}$ of item $I_j$ is updated by Eqn. (\ref{eqn:message_aggregate_i}).
\begin{equation}
   \hat{\mathbf{C}}^{(l)}_{I_j}=\sum_{u_{k, g}\in\mathcal{N}^{I_{j}}_{u}}\mathbf{m}^{(l)}_{I_{j}\leftarrow u_{k,g}} + \sum_{I_{j-1}\in\mathcal{N}^{I_{j}}_{I}}\mathbf{m}^{(l)}_{I_{j}\leftarrow I_{j-1}}, \label{eqn:message_aggregate_i}
\end{equation}
where $\mathcal{N}^{I_j}_{u}$ is the set of all latent users who have interactions on item $I_j$, and $\mathcal{N}^{I_j}_{I}$ denotes the set of all neighboring items of $I_j$.

By adopting layer-wise message aggregation, the final representations of $I_j$ and $u_{k, h}$ are denoted as follows:
\begin{equation}
    \mathbf{E}_{I_j} = \sum_{l=0}^{L}\hat{\mathbf{C}}_{I_j}^{(l)}; \quad \mathbf{E}_{u_{k,h}} = \sum_{l=0}^{L} \hat{\mathbf{C}}_{u_{k,h}}^{(l)}, 
\end{equation}
where $L$ is a hyper-parameter that controls the layer number of the graph convolutions on the primary capsule graphs.
\subsubsection{Account-level Dynamic Routing.}
The account-level dynamic routing mechanism performs a routing selection to consider the associations between account and its latent users. The strength of the connections between user capsules and account capsules are qualified via a coupling coefficient $b_{p}$, which is initialized randomly. The dynamic routing is operated iteratively for $\theta$ times, where $\theta$ is a hyper-parameter. As a common practice \cite{Zheng_Capsule_Graph_22_SIGIR}, we uniformly set $\theta$ as 3, maintaining a balance between performance and computational complexity. The dynamic routing at the $j$-th iteration is represented as Eqn. (\ref{eqn:dynamic_routing}).
\begin{equation}
\tilde{\mathbf{C}}_{A_k}^{(j)} = \sum_{h}^{A_k}\text{squash}\left(\sum_{p} b_{p}^{(j-1)} \mathbf{E}_{u_{k,h}}\right), \label{eqn:dynamic_routing}
\end{equation}
where $\tilde{\mathbf{C}}_{A_k}^{(j)}$ denotes the account capsule for $A_k$, $\text{squash}(\cdot)$ is the squash function which compresses the routed information, ensuring efficient information transmission.

In addition, the coupling coefficient $b_{p}$ at the $j$-th iteration is updated by calculating the affinity between user capsules and the account capsule:
\begin{equation}
b_{p}^{(j)} = b_{p}^{(j-1)} + \mathbf{W}_d\sum_{h}^{A_k}\left(\mathbf{E}_{u_{k,h}} \odot \tilde{\mathbf{C}}_{A_k}^{(j)}\right), \label{eqn:update_ef}
\end{equation}
where $\mathbf{W}_d$ is a learnable weighting matrix, $\odot$ denotes the element-wise product.

Hence, the final representations of accounts and sequences are denoted as Eqn. (\ref{eqn:concat_cap}).
\begin{equation}
    \mathbf{E}_A = \sum_{A_k \in \mathcal{A}} \tilde{\mathbf{C}}_{A_k}^{(\theta)}; \quad \mathbf{E}_S = \sum_{S_k \in \mathcal{S}} \sum_{I_j \in S_k} \mathbf{E}_{I_j}. \label{eqn:concat_cap}
\end{equation}

With the help of the Graph Capsule Convolutional Network (GC$^2$N), the account representation has gained the ability to finely distinguish the preferences of various potential users within shared accounts. However, the diverse preferences within sequence representations remain largely unexplored. Hence, we further devise a subspace alignment method.

\subsection{Subspace Alignment (SA)}
Subspace alignment is an efficient component that clusters the hybrid preferences of multiple latent users and then aligns them to the sequence representations $\mathbf{E}_S$. 
Instead of using traditional self-representation matrices \cite{Xie_Sub_22_ICDE, Zhang_Sub_2018_WISE}, SA exploits low-rank subspace bases to cluster diverse preferences for various latent users. Moreover, it also refines the sequence representations by attaching the subspace affinities, and then aligns them with the original sequence representations via a contrastive learning strategy. This strategy provides additional self-supervised signals to distinguish the preferences of latent users within hybrid sequences.

\subsubsection{Subspace Affinity Calculation.} Taking the sequence $S_k \in \mathcal{S}$ as an example, $\mathbf{E}_{S_k} \in \mathbb{R}^{n \times d_2}$ denotes its representations. The initialization of subspace bases $\mathbf{D}=\{\mathbf{d}_1, \mathbf{d}_2, \dots, \mathbf{d}_j, \dots, \mathbf{d}_{\alpha}\} \in \mathbb{R}^{\alpha \times d_2}$ is given by the column space of the clusters generated by K-means on $\mathbf{E}_{S_k}$. Then, the subspace affinities are calculated as:
\begin{equation}
    s_{ij}=\frac{\left\|\mathbf{e}_i^\top \mathbf{d}_j\right\|_F^2+\lambda d_2}{\sum_j\left(\left\|\mathbf{e}_i^\top\mathbf{d}_j\right\|_F^2+\lambda d_2\right)}, \label{eqn:sub_aff}
\end{equation}
where $s_{ij} $ denotes the subspace affinity between $i$-th item $\mathbf{e}_i$ in $S_k$ and $j$-th subspace base $\mathbf{d}_j$, $\lambda$ is a parameter that controls the smoothness of the calculation ($\lambda$ is uniformly set to 1e-4 according to the common practice reported in \cite{Cai_subspace_22_CVPR}).

As the affinity calculation does not rely on the self-expression framework, SA is able to achieve linear computational complexity (i.e., O($N \times d^2$)) and low memory consumption, making subspace clustering more efficient. 
\subsubsection{Contrastive Learning.} 
To align the sequence embeddings with the clustered user preferences, SA first refines the $\mathbf{e}_i \in \mathbb{R}^{1 \times d_2}$ by applying the affinities:
\begin{equation}
    \mathbf{z}_i = \mathbf{e}_i \cdot s_{ij}, \label{eqn:refine_seq}
\end{equation}
where $\mathbf{z}_i \in \mathbf{Z}_{S_k}$ denotes the refined embedding of $i$-th item in $S_k$ and $\mathbf{Z}_{S_k} \in \mathbb{R}^{n \times d_2}$ represents the refined representation of $S_k$.

Then, SA adopts a contrastive learning paradigm that aligns the refined representations $\mathbf{Z}_{S_k}$ with the normalized sequence embeddings $\mathbf{\hat{E}}_{S_k}$:
\begin{equation}
    \mathcal{L}_{C}^{S_k}=-\sum_{\mathbf{z}_i \in \mathbf{Z}_{S_k}}^{\mathbf{\hat{e}}_i \in \mathbf{\hat{E}}_{S_k}}\log\left(\frac{\text{exp}(\mathbf{\hat{e}}_{i} \odot \mathbf{z}_{i}/\beta)}{\sum_{j\in \mathcal{I}} \text{exp}(\mathbf{\hat{e}}_{i} \odot \mathbf{z}_{j}/\beta)}\right), \label{eqn:ssl_sub}
\end{equation}
where $\mathcal{L}_{C}^{S_k}$ is the InfoNCE loss calculated between representations of $S_k$, ($\mathbf{\hat{e}}_i$, $\mathbf{z}_i$) is a positive pair while ($\mathbf{\hat{e}}_i$, $\mathbf{z}_j$) represents a negative pair, $\beta$ denotes the temperature coefficient that controls the impact from negative pairs to positive pairs. 

The contrastive loss for all the sequences is:
\begin{equation}
    \mathcal{L}_{C} = \sum_{S_k \in \mathcal{S}} \mathcal{L}_{C}^{S_k}. \label{eqn:ssl_all}
\end{equation}

Such a strategy is able to distinguish various preferences of latent users. Therefore, it improves the capability of learning fine-grained sequence representations. The final sequence embeddings are obtained by summing the refined sequence embeddings and their normalized-forms:
\begin{equation}
    \mathbf{\hat{E}}_S = \sum_{S_k \in \mathcal{S}} \text{Norm} \left( \mathbf{\hat{E}}_{S_k} + \mathbf{W}_s \mathbf{Z}_{S_k} \right), \label{eqn:sum_sequence}
\end{equation}
where $\mathbf{\hat{E}}_S$ denotes the updated sequence embeddings, $\mathbf{W}_s$ is the weighting matrix.

\subsection{Final Prediction}
The final prediction generated by LightGC$^2$N is denoted as:
\begin{equation}
    P(I_{t+1}|\mathcal{S}, \mathcal{A}) = \text{softmax}\left( \mathbf{W}_f \cdot [\mathbf{\hat{E}}_S, \mathbf{E}_A]^\top + \mathbf{b}_f
 \right),
\end{equation}
where $\mathbf{W}_f$ is the transformation matrix that maps the predictions to the dimension of candidate items, and $\mathbf{b}_f$ is the bias term that adjusts the threshold of the activation function.

The cross-entropy loss function is adopted to optimize the learnable parameters in LightGC$^2$N, which is denoted as:
\begin{equation}
    \mathcal{L}_S = -\frac{1}{|\mathcal{S}|}\sum_{I_{t+1}\in\mathcal{I}}\text{log} P(I_{t+1}|\mathcal{S}, \mathcal{A}).
\end{equation}

Then, the overall loss function is denoted as:
\begin{equation}
    \mathcal{L} = \mathcal{L}_S + \gamma \mathcal{L}_C,
\end{equation}
where $\gamma$ is a hyper-parameter that controls the participation of self-supervised signals.

\section{Experiments}
In this section, we first introduce the experimental settings, and then analyze the performance of LightGC$^2$N by answering the following \textbf{R}esearch \textbf{Q}uestions. 
% \textbf{RQ1:} How does the performance of LightGC$^2$N in terms of training efficiency and parameter scale? \textbf{RQ2:} How does LightGC$^2$N perform on the SSR compared with other state-of-the-art methods? \textbf{RQ3:} How do the key components of LightGC$^2$N contribute to the recommendation performance? \textbf{RQ4:} How do the hyper-parameters affect the performance of LightGC$^2$N?

\begin{itemize}
    \item \textbf{RQ1:} How does the performance of LightGC$^2$N in terms of training efficiency and parameter scale?
    \item \textbf{RQ2:} How does LightGC$^2$N perform on the SSR compared with other state-of-the-art methods? 
    \item \textbf{RQ3:} How do the key components of LightGC$^2$N contribute to the recommendation performance?
    \item \textbf{RQ4:} How do the hyper-parameters affect the performance of LightGC$^2$N?
\end{itemize}
\subsection{Experimental Settings}
\subsubsection{Datasets.} We evaluate LightGC$^2$N on four real-world datasets released by \cite{Ma_Shared_19_SIGIR}, including Hvideo-E (HV-E), Hvideo-V (HV-V), Hamazon-M (HA-M), and Hamazon-B (HA-B). HV-E and HV-V are two smart TV datasets comprising viewing logs from different TV channels. HV-E encompasses logs of educational videos and instructional content in areas such as sports nutrition and medicine, whereas HV-V includes logs of television series and films. HA-M and HA-B are derived from two Amazon domains, featuring movie viewing (HA-M) and book reading (HA-B). For the evaluation, we randomly assigned 80$\%$ of the sequences to the training set, and the remaining 20$\%$ to the testing set. Note that, the most recently observed item in each sequence per dataset is designated as the ground truth item.
\begin{table*}[h]
  \centering
  \small
      \setlength{\tabcolsep}{1.45mm}
    \begin{tabular}{l|c|cc|ccc|cccc|c}
    \toprule
    \midrule
    \multicolumn{1}{c|}{\textbf{Dataset}} & \multicolumn{1}{c|} {\textbf{Metric}} & \multicolumn{1}{c}{NCF} & \multicolumn{1}{c|}{LightGCN} & \multicolumn{1}{c}{HRNN} & \multicolumn{1}{c}{NAIS} & \multicolumn{1}{c|}{TGSRec} & \multicolumn{1}{c}{$\pi$-Net} & \multicolumn{1}{c}{PSJNet} & \multicolumn{1}{c}{DA-GCN} & \multicolumn{1}{c|}{TiDA-GCN} & \multicolumn{1}{c}{\textbf{LightGC$^2$N}}\\
    \midrule
    \multicolumn{1}{c|}{\multirow{4}[1]{*}{\textbf{HV-E}}}& \multicolumn{1}{c|}{RC@5} & 11.25& 20.70& 22.55& 19.80& 19.91& 25.13& 24.80& 51.35 & \underline{54.11}& \textbf{61.35}\\
    &  \multicolumn{1}{c|}{RC@20} & 20.12& 39.92& 47.98& 40.17& 41.80& 47.08& 46.68&66.93 & \underline{68.98}& \textbf{72.73}\\
    &  \multicolumn{1}{c|}{MRR@5} & 5.77& 11.55& 13.76& 11.45& 13.95& 15.36& 15.37&35.63 & \underline{38.66}& \textbf{46.24}\\
    &  \multicolumn{1}{c|}{MRR@20} & 7.85& 13.56& 16.14& 13.24& 15.73& 17.52& 17.56&37.27 & \underline{40.23}& \textbf{47.35}\\
    \midrule
    \multicolumn{1}{c|}{\multirow{4}[1]{*}{\textbf{HV-V}}}& \multicolumn{1}{c|}{RC@5} & 27.21& 58.45& 68.00& 59.30& 58.91& 67.00& 66.86&75.39 & \underline{76.37}& \textbf{79.29}\\
    &  \multicolumn{1}{c|}{RC@20} & 22.63& 63.55& 73.24& 67.41& 67.22& 74.17& 74.14&82.37 & \underline{83.58}& \textbf{84.70}\\
    &  \multicolumn{1}{c|}{MRR@5} & 22.99& 54.00& 60.58& 51.52& 50.32& 60.37& 61.89&59.78 & \underline{63.58}& \textbf{66.27}\\
    &  \multicolumn{1}{c|}{MRR@20} & 24.71& 56.72& 63.31& 54.47& 53.40& 61.74& 62.63&60.55 & \underline{65.37}& \textbf{66.87}\\
    \midrule
    \multicolumn{1}{c|}{\multirow{4}[1]{*}{\textbf{HA-M}}}& \multicolumn{1}{c|}{RC@5} &7.82 &15.54 &16.96 & 14.03& 14.59&18.54 &16.25 &22.93 &\underline{23.55} & \textbf{46.02}\\
    &  \multicolumn{1}{c|}{RC@20} &10.34 &18.20 &20.81 & 16.02& 18.44&21.87 &18.14 &23.90 &\underline{24.33} & \textbf{48.46}\\
    &  \multicolumn{1}{c|}{MRR@5} &2.72 &12.88 &13.75 & 10.55& 11.91&16.24 &11.25 &20.09 &\underline{20.91} & \textbf{41.30}\\
    &  \multicolumn{1}{c|}{MRR@20} &3.11 &13.12 &14.14 & 12.57& 14.00&16.56 &13.58 &20.19 &\underline{21.23} & \textbf{41.64}\\
    \midrule
    \multicolumn{1}{c|}{\multirow{4}[1]{*}{\textbf{HA-B}}}& \multicolumn{1}{c|}{RC@5} &8.31 &21.14 &20.92 & 14.51& 14.57&22.44 &16.67 &23.93 &\underline{24.69} & \textbf{47.36}\\
    &  \multicolumn{1}{c|}{RC@20} &11.22 &22.88 &23.64 & 19.82& 19.93&23.75 &19.30 &24.25 &\underline{24.82} & \textbf{48.55}\\
    &  \multicolumn{1}{c|}{MRR@5} &7.92 &15.58 &17.04 & 13.29& 15.74&20.38 &15.52 &21.35 &\underline{21.88} & \textbf{44.40}\\
    &  \multicolumn{1}{c|}{MRR@20} &9.88 &17.30 &17.35 & 15.99& 16.63&20.58 &17.30 &21.39 &\underline{22.21} & \textbf{44.52}\\
    \midrule
    \bottomrule
    \end{tabular}
    \caption{Experimental results ($\%$) for different methods on four real-world datasets. The best results are indicated in bold, while underlined values indicate the sub-optimal results.}
  \label{tab:experiments}
\end{table*}

\subsubsection{Evaluation Metrics.} For model evaluations, we adopt two common evaluation metrics \cite{Guo_Shared_21_IJCAI} to assess the model performance, i.e., top-$N$ Recall (Recall@N) and top-$N$ Mean Reciprocal Rank (MRR@N), where $N=\{5, 20\}$.

\subsubsection{Implementation Details.} 
We implemented LightGC$^2$N with TensorFlow and accelerated the model training using an Intel$^\circledR$ Xeon$^\circledR$ Silver 4210 CPU (2.20GHz) and NVIDIA$^\circledR$ RTX 3090 (24G) GPU. The operating system is Ubuntu 22.04, the system memory is 126G, and the coding platform is PyCharm. The learnable parameters are initialized via Xavier \cite{Xavier_experiment_2010_AISTATS}, the loss function is optimized by Adam \cite{Adam_experiment_2015_ICLR} optimizer. For the training settings, we set the batch-size as 256, the learning rate as 0.005, the dropout rate as 0.1, and the training epochs as 200. We uniformly set the embedding-size as 16 for LightGC$^2$N and other baseline methods to ensure the fairness of experiments. For other hyper-parameters of baselines, we adopt optimal hyper-parameter settings reported in their paper and then fine-tuned them on each dataset. 

\subsubsection{Baselines.}
To validate the performance of LightGC$^2$N on SSR, we compared it with the following baselines: 
1) Traditional recommendations: NCF \cite{He_baseline_17_WWW}, and LightGCN \cite{He_Graph_20_SIGIR}. 2) Sequential recommendations:  HRNN \cite{Quadrana_SR_17_RecSys}. NAIS \cite{He_SR_18_TKDE}, and TGSRec \cite{Fan_SR_21_CIKM}.
3) Shared-account sequential recommendations: $\pi$-Net \cite{Ma_Shared_19_SIGIR}, PSJNet \cite{Sun_Shared_23_TKDE}, DA-GCN \cite{Guo_Shared_21_IJCAI}, and TiDA-GCN \cite{Guo_Shared_24_TNNLs}.

\subsection{Parameter Scale and Training Efficiency (RQ1)}
\begin{figure}[ht]
\centering
\includegraphics[width=0.85\columnwidth]{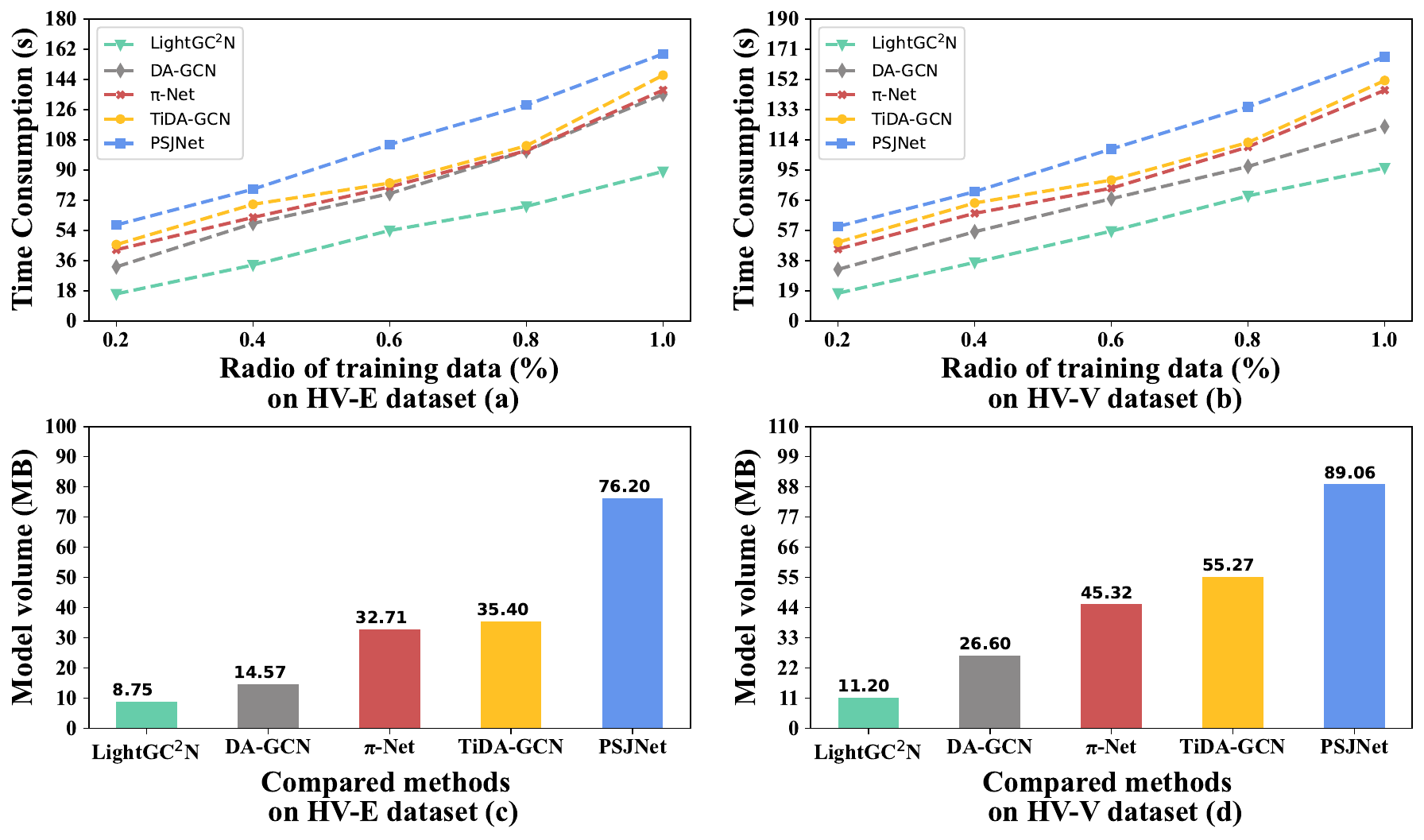} % Reduce the figure size so that it is slightly narrower than the column. Don't use precise values for figure width.This setup will avoid overfull boxes.
\caption{Comparison of time consumption and parameter scale between LightGC$^2$N and competitive SSR methods.}
\label{fig3:efficiency}
\end{figure}

In this section, we initially vary the proportion of input data from 0.2 to 1.0 on the HV-E and HV-V datasets to assess the LightGC$^2$N’s training time consumption. Subsequently, we compare its parameter scale with other competitive methods, i.e., PSJNet, TiDA-GCN, $\pi$-net and DA-GCN. The observations are as follows: 1) Figure \ref{fig3:efficiency} (a) and (b) reveal that LightGC$^2$N exhibits reduced training time compared to other baselines, signifying enhanced training efficiency and scalability for large-scale datasets. 2) As depicted in Figure \ref{fig3:efficiency} (c) and (d), LightGC$^2$N requires notably fewer parameters than other methods, providing a positive answer to RQ1.
\subsection{Overall Performance (RQ2)}
Table \ref{tab:experiments} shows the experimental results of LightGC$^2$N compared with other state-of-the-art methods on four datasets. The observations are summarized as follows: 1) The sequential recommendation methods (i.e., HRNN, NAIS, and TGSRec) perform better than traditional methods (i.e., NCF and LightGCN). This observation indicates the significance of modeling users' sequential preferences. 2) The SSR solutions (i.e., $\pi$-Net, PSJNet, DA-GCN, TiDA-GCN, and LightGC$^2$N) typically outperform the other traditional and sequential recommendation methods, demonstrating the significance of addressing the shared-account issues in real-world sequential recommendation scenarios. 3) LightGC$^2$N outperforms other state-of-the-art SSR methods (i.e., $\pi$-Net, PSJNet, DA-GCN, and TiDA-GCN). This observation indicates the significance of capturing the fine-grained differences among latent users for SSR. 4) LightGC$^2$N exceeds other graph-based SSR methods (i.e., DA-GCN and TiDA-GCN), demonstrating the superiority of our proposed graph capsule convolutional networks in modeling complicated associations for SSR. 5) LightGC$^2$N achieves the best performance on all datasets, demonstrating the superiority of our proposed graph capsule convolutional network and subspace alignment method for the SSR scenarios.
\begin{table}[h]
  \centering
  \small

   \setlength{\tabcolsep}{0.6mm}
    \begin{tabular}{l|cccc|cccc}
    \toprule
    \midrule
    \multicolumn{1}{c|}{\textbf{Dataset}} & \multicolumn{4}{c|}{\textbf{HV-E}} &
    \multicolumn{4}{c}{\textbf{HV-V}} \\
    \midrule
    \multicolumn{1}{c|}{\multirow{2}[1]{*}{\textbf{Metric}}}& \multicolumn{2}{c}{\textbf{Recall}} & \multicolumn{2}{c|}{\textbf{MRR}}& \multicolumn{2}{c}{\textbf{Recall}} & \multicolumn{2}{c}{\textbf{MRR}} \\
    \cmidrule{2-9}
    & \textbf{@5} & \textbf{@20} & \textbf{@5} & \textbf{@20}
    & \textbf{@5} & \textbf{@20} & \textbf{@5} & \textbf{@20}
          \\
    \midrule
    Light$_{w/o\text{LA}}$ &58.33 &67.58 &39.71 &40.24
    &76.01 &81.61 &63.17 &64.04\\
    Light$_{w/o\text{DR}}$ &59.86 &68.66 &40.52 &41.44
    &77.33 &82.15 &64.22 &64.75\\
    Light$_{w/o\text{C}}$ &53.46 &66.88 &35.42 &36.91
    &73.12 &78.88 &61.72 &63.92\\
    Light$_{w/o\text{CL}}$ &59.96 &71.12 &43.51 &44.68
    &77.02 &83.15 &64.85 &65.10\\
    Light$_{w/o\text{S}}$ &57.96 &70.12 &40.51 &41.68
    &75.12 &82.16 &62.88 &64.05\\
    Light$_{w/o\text{A}}$  &22.90 &43.42 &16.26 &22.97
    &60.58 &65.30 &54.94 &57.82\\
    \midrule
    \textbf{LightGC$^2$N} &\textbf{61.35} &\textbf{72.73} &\textbf{46.24} &\textbf{47.35}
    &\textbf{79.29}&\textbf{84.70}&\textbf{66.27}&\textbf{66.87}\\
    \midrule
    \bottomrule
    \end{tabular}
   \caption{The experimental results $(\%)$ of ablation studies on two real-world datasets.}
  \label{tab:ablation_studies}
\end{table}
\subsection{Ablation Study (RQ3)}
In this section, we conduct a series of ablation studies on HV-E and HV-V to explore the impact of different components for LightGC$^2$N. As shown in Table \ref{tab:ablation_studies}, 1) Light$_{w/o\text{C}}$ is a variant that replace Graph Capsule Convolutional Network (GC$^2$N) by traditional graph convolutional network. 2) Light$_{w/o\text{S}}$ is another variant method that disables the Subspace Alignment (SA) component. 3) Light$_{w/o\text{LA}}$ is a variant that replaces Linear attention with Conv1D when projecting item embeddings into capsule-form. 4) Light$_{w/o\text{DR}}$ is another variant that omits the account-level dynamic routing in GC$^2$N. 5) Light$_{w/o\text{CL}}$ is a method that excludes the contrastive learning in SA. 6) Light$_{w/o\text{A}}$ is a variant that removes all the component of LightGC$^2$N. 

The observations of Table \ref{tab:ablation_studies} are summarized as follows: 1) LightGC$^2$N outperforms Light$_{w/o\text{C}}$ and Light$_{w/o\text{A}}$. This observation demonstrates that GC$^2$N works well in distinguishing preferences of latent users. It also demonstrates that the fine-grained distinction of the preferences for different latent users indeed enhance the model performance on SSR. 2) LightGC$^2$N outperforms Light$_{w/o\text{LA}}$, illustrating the effectiveness of Linear attention in capturing global correlations among items. 3) LightGC$^2$N outperforms Light$_{w/o\text{DR}}$, suggesting that the routing selection in the account-level dynamic routing indeed improve the performance of GC$^2$N. It also demonstrates that the account-level dynamic routing component performs well in merging the preferences from multiple latent users. 4) LightGC$^2$N performs better than Light$_{w/o\text{S}}$, demonstrating the contribution of the subspace alignment method for the sequence-level representation learning. 5) LightGC$^2$N outperforms Light$_{w/o\text{CL}}$, indicating that the contrastive learning strategy aids in subspace alignment between refined and original sequence embeddings.
\begin{figure}[t]
    \centering
    \begin{minipage}[t]{1\linewidth}
        \centering
        \includegraphics[width=0.9\columnwidth]{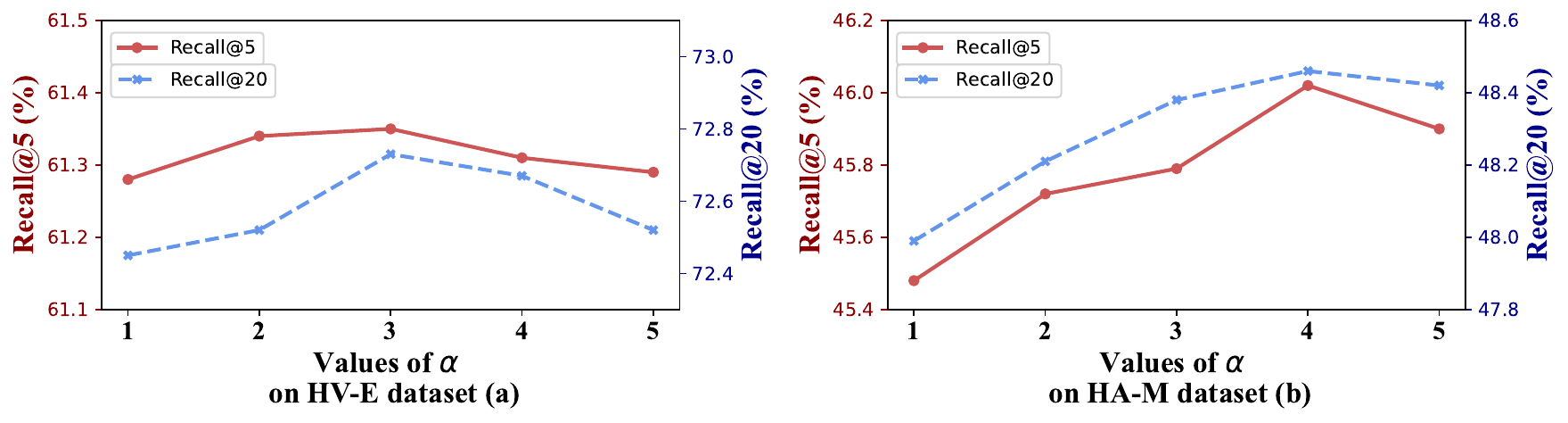}
    \end{minipage}
    \begin{minipage}[t]{1\linewidth}
        \centering
        \includegraphics[width=0.9\columnwidth]{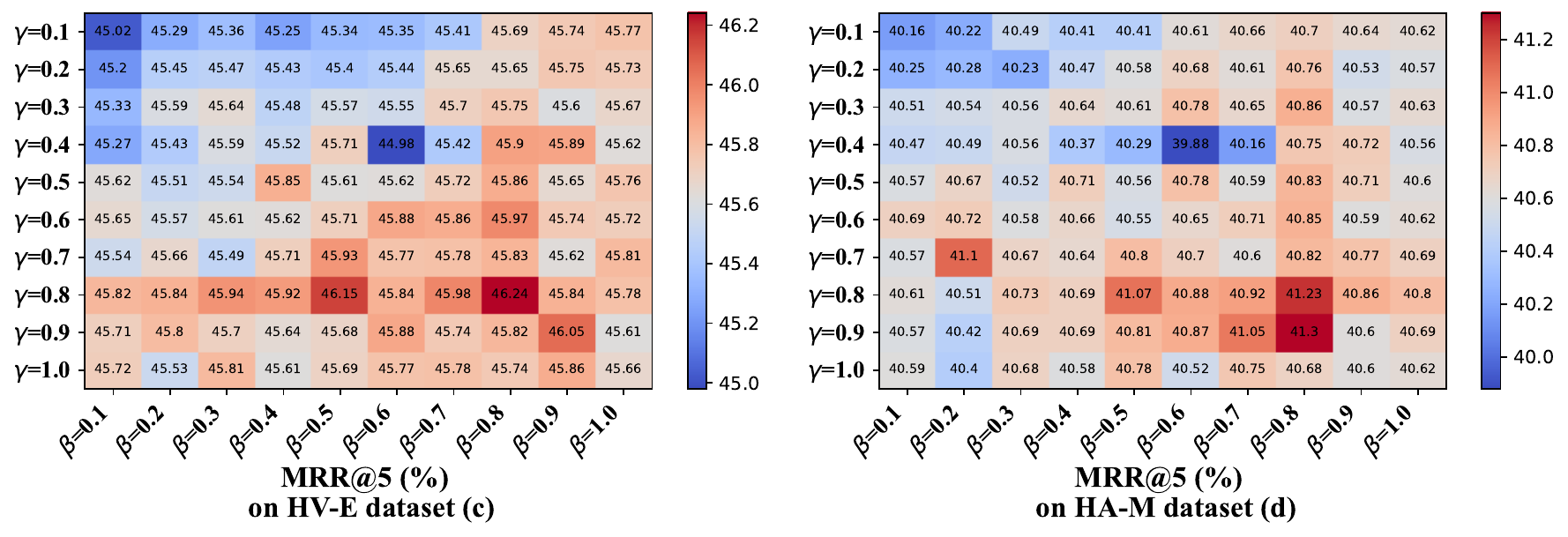}
    \end{minipage}
    \caption{Impact of hyper-parameters $\alpha$, $\beta$ and $\gamma$ on HV-E and HA-M.}
    \label{fig:hyper_param}
\end{figure}
\subsection{Hyper-parameters Analysis (RQ4)}
The hyper-parameter $\alpha$ controls the number of latent users within each shared account. Figures \ref{fig:hyper_param} (a) and (b) show the performance of LightGC$^2$N with different $\alpha$ values in $\{1, 2, 3, 4, 5\}$ on two different datasets (HV-E and HA-M). The experimental results indicate that LightGC$^2$N requires different parameter settings to achieve optimal performance on different datasets, which is consistent with the real-world setting (i.e., the number of latent users sharing an account varies in different scenarios). $\beta$ and $\gamma$ are two significant hyper-parameters that respectively control the temperature and the participation of the contrastive learning. As shown in Figures \ref{fig:hyper_param} (c) and (d), LightGC$^2$N reach the best performance when they are set to 0.8 or 0.9. This observation further validates the effectiveness of our contrastive learning strategy. Additionally, it indicates that the self-supervised signals contribute to sequence-level representation learning only when these hyper-parameters are appropriately valued. 
\section{Conclusion and Future Work}
In this work, we introduce a Lightweight Graph Capsule Convolutional Network (LightGC$^2$N) with subspace alignment to tackle the problems in Shared-account Sequential Recommendation (SSR). By effectively capturing the fine-grained preferences of latent users, LightGC$^2$N achieves the best performance on various datasets. The lightweight design of this work makes it suitable for deployment on resource-constrained devices while maintaining high recommendation accuracy. Experimental results on four SSR datasets demonstrates the effectiveness and efficiency of LightGC$^2$N, paving the way for its practical application in real-world recommendation systems.

However, LightGC$^2$N assumes a fixed number of latent users within each shared account. In our future work, we will study how to determine the number of latent users in each account automatically or heuristically. 

% \section*{Acknowledgments}
% This research is supported by the National Natural Science Foundation of China (Grant No. 62472263, 62072288), the Taishan Scholar Program of Shandong Province, Shandong Youth Innovation Team, the Natural Science Foundation of Shandong Province (Grant No. ZR2024MF034, ZR2022MF268).

\bibliography{aaai25}

\newpage

\appendix

\section{Technical Appendix}
\subsection{A1. Notations}
The key notations of embedded vectors and matrices that utilized in the Methodology are summarized in Table \ref{tab:notations}.
\begin{table}[h]
\centering
\caption{The notations mainly used in this paper.}
\label{tab:notations}
\footnotesize
\begin{tabular}
{p{0.2\columnwidth}|p{0.68\columnwidth}}%{@{}ll@{}}
\toprule
\midrule
\textbf{Notations} &  \textbf{Descriptions} \\ 
\midrule
$\mathbf{E}_I^{(0)}$ & the node embeddings of items at 0-th layer \\
$\mathbf{E}_A^{(0)}$ & the node embeddings of accounts at 0-th layer\\
$\mathbf{C}_I^{(0)}$ & the capsule embeddings of items at 0-th layer\\
$\mathbf{C}_A^{(0)}$ & the capsule embeddings of accounts at 0-th layer\\
$\mathbf{C}_{u_{k, h}}$ & the capsule embedding of $h$-th latent user in shared-account $A_k$\\
$\mathbf{C}_{I_j}$ & the capsule embedding of $j$-th neighbor item for latent user $u_{k, h}$\\
$\mathbf{m}_{u_{k,h}\leftarrow I_{j}}$ & the messages passed from neighbor items to user $\mathbf{C}_{u_{k, h}}$\\
$\mathbf{m}_{u_{k,h}\leftarrow u_{k,h}}$ & the self-connection messages of $\mathbf{C}_{u_{k, h}}$\\
$\mathbf{m}_{I_{j}\leftarrow u_{k,g}}$ & the messages passed from neighbor users to item $\mathbf{C}_{I_j}$\\
$\mathbf{m}_{I_{j}\leftarrow I_{j-1}}$ & the messages passed from neighbor items to item $\mathbf{C}_{I_j}$\\
$\tilde{\mathbf{C}}_{A_k}$ & the resulted capsule embeddings of shared accounts by dynamic routing\\
$\mathbf{E}_A$ & the output embeddings of accounts from GC$^2$N\\
$\mathbf{E}_S$ & the output embeddings of sequences from GC$^2$N\\
$\mathbf{Z}_{S_k}$ & the refined embedding of sequence $S_k$ by applying subspace affinities\\
$\mathbf{\hat{E}}_{S_k}$ & the normalized embedding of $S_k$ for contrastive learning\\
$\mathbf{\hat{E}}_S$ & the output embeddings of sequences from SA\\
\midrule
\bottomrule
\end{tabular}
\end{table}
\subsection{A2. Detailed Dataset Description}
We evaluate LightGC$^2$N on four SSR-oriented datasets (i.e., HV-E, HV-V, HA-M and HA-B), which are released by \cite{Ma_Shared_19_SIGIR}. The statistics of all datasets are presented in Table \ref{tab:dataset_statistics}. 
\begin{table}[h]
    \centering
    \small
     \caption{Statistics of four real-world datasets.}
    \setlength{\tabcolsep}{1mm}
    \begin{tabular}{l|c|c|c|c}
    \toprule
    \midrule
    \multicolumn{1}{c|}{\textbf{Dataset}}&\multicolumn{1}{c|}{\textbf{HV-E}}&\multicolumn{1}{c|}{\textbf{HV-V}}&\multicolumn{1}{c|}{\textbf{HA-M}}&\multicolumn{1}{c}{\textbf{HA-B}}\\
    \midrule
    Items &8,367 &11,404 &67,161 &126,547\\
    Interactions &2,129,500 &1,893,784 &2,196,574 &2,135,995\\
    \midrule
    Accounts & \multicolumn{2}{c|}{13,714} &\multicolumn{2}{c}{13,724}\\
    Training Seqs. &\multicolumn{2}{c|}{114,197} &\multicolumn{2}{c}{122,303}\\
    Testing Seqs. &\multicolumn{2}{c|}{20,152} &\multicolumn{2}{c}{21,582}\\
    \midrule
    \bottomrule
    \end{tabular}
    \label{tab:dataset_statistics}
\end{table}

HV-E and HV-V are two datasets sourced from a smart TV platform which has 13,714 accounts and their 134,349 hybrid sequences. These two datasets are well-suited for SSR due to the shared nature of family accounts \cite{Guo_Shared_24_TNNLs}, which contains family account viewing logs in the education (E-domain) and video-on-demand (V-domain) from October 2016 to June 2017. 

HA-M and HA-B are two real-world datasets of user reviews from Amazon, encompassing the movie (M-domain) and book (B-domain) categories. These datasets span from May 1996 to July 2014 and include 13,724 unique accounts and 143,885 hybrid sequences. However, they are not originally tailored for SSR. To simulate shared accounts, \cite{Ma_Shared_19_SIGIR} merged 2–4 users into shared accounts to generate the hybrid sequences. Each sequence is divided into small fragments by year, and those sequences with insufficient items (less than 5 interactions) in either domain were excluded. 
\subsection{A3. Detailed Baseline Settings}

\textbf{1) Traditional recommendations:} To ensure the fairness of the experimental results, we feed the these traditional methods with sequential inputs. 
\begin{itemize}
    \item \textbf{NCF} \cite{He_baseline_17_WWW}: This is a traditional recommendation method that exploits deep neural networks to capture the collaborative filtering between interactions.
    \item \textbf{LightGCN} \cite{He_Graph_20_SIGIR}: LightGCN is a simplified graph-based method for traditional recommendation 
\end{itemize}
\textbf{2) Sequential recommendations:} 
\begin{itemize}
    \item \textbf{HRNN} \cite{Quadrana_SR_17_RecSys}: HRNN is an early proposed RNN-based SR method, which devises a hierarchical GRU structure to learn the sequential representations.
    \item \textbf{NAIS} \cite{He_SR_18_TKDE}: This is an attention-based SR method that designs a nonlinear attention network to calculate the correlations among items.
    \item \textbf{TGSRec} \cite{Fan_SR_21_CIKM}: This is a time interval-aware SR method, which considers the time gaps among interactions via an attention-based network.
\end{itemize}

\textbf{3) Shared-account sequential recommendations:} 
\begin{itemize}
    \item \textbf{$\pi$-Net} \cite{Ma_Shared_19_SIGIR}: This is an RNN-based SSR method that transfers knowledge between domains and models the shared-account preferences via the Gating Recurrent Units (GRUs).
    \item \textbf{PSJNet} \cite{Sun_Shared_23_TKDE}: This is another RNN-based SSR method that further improves the $\pi$-Net via a hierarchical split and joint strategy.
    \item \textbf{DA-GCN} \cite{Guo_Shared_21_IJCAI}: DA-GCN is a graph-based SSR method that leverages an attention enhance graph convolutional networks to identify the preferences of various latent users and share the information between different domains.
    \item \textbf{TiDA-GCN} \cite{Guo_Shared_24_TNNLs}: This is a state-of-the art graph-based SSR method that further improve DA-GCN by considering time intervals between interactions.
\end{itemize}   

As all the competitive SSR methods are cross-domain methods, we refer and report their performance under the cross-domain settings in Table \ref{tab:experiments}. In another word, we allow them transferring knowledge between domains (i.e., between HV-E and HV-V, or between HA-M and HA-B) to reach their best performance.
\subsection{A4. Algorithm Details}
Algorithm \ref{alg:gc2n} shows the pseudo-codes for the Graph Capsule Convolutional Network (GC$^2$N) which is one of the key component in LightGC$^2$N.
\begin{algorithm}
\caption{Graph Capsule Convolutional Network}
\label{alg:gc2n}
\textbf{Input}: Input Laplace matrix $\mathcal{M}$, item embeddings $\mathbf{E}_I^{(0)}$, account embeddings $\mathbf{E}_A^{(0)}$

\textbf{Output}: Output the sequence embeddings $\mathbf{E}_{S}$ and the account embeddings $\mathbf{E}_{A}$
\begin{algorithmic}[1] %[1] enables line numbers
\STATE Construct primary capsule graphs $\mathcal{G}_{c}$:
\STATE $\quad$ Utilize linear attention to project item embeddings $\mathbf{E}_I^{(0)}$ into capsules $\mathbf{C}_I^{(0)}$ via Eqn. (\ref{eqn:LA}).
\STATE $\quad$ Utilize Conv1D to project account embeddings $\mathbf{E}_A^{(0)}$ into capsules $\mathbf{C}_A^{(0)}$ via Eqn. (\ref{eqn:conv1d}).
\STATE $\quad$ Split $\mathbf{C}_A^{(0)}$ into $\alpha$ user capsules $\mathbf{C}_u^{(0)}$.
\STATE Message passing on primary capsule graph:
\FOR{$l = 0 \rightarrow L$} 
    \STATE For each latent user capsule $\mathbf{C}^{(l)}_{u_{k, h}}$:
        \STATE $\quad$ Calculate messages passed from neighbor item capsules $\mathbf{m}^{(l)}_{u_{k,h}\leftarrow I_{j}}$ by Eqn. (\ref{eqn:message_iu}).
        \STATE $\quad$ Calculate self-connection messages of the user capsule $\mathbf{m}^{(l)}_{u_{k,h}\leftarrow u_{k,h}}$ by Eqn.(\ref{eqn:meesage_uu}).
        \STATE $\quad $ Aggregate the above messages. 
    \STATE For each item capsule $\mathbf{C}^{(l)}_{I_j}$:
        \STATE $\quad$ Calculate messages passed from neighbor user capsules $\mathbf{m}^{(l)}_{I_{j}\leftarrow u_{k,g}}$ by Eqn. (\ref{eqn:meesage_i}-1).
        \STATE $\quad$ Calculate messages passed from neighbor item capsules $\mathbf{m}^{(l)}_{I_{j}\leftarrow I_{j-1}}$ by Eqn. (\ref{eqn:meesage_i}-2).
        \STATE $\quad$ Aggregate the above messages. 
    \STATE Update the representations of user $\mathbf{C}_{u_{k, h}}^{(l)}$ and item $\mathbf{C}_{I_j}^{(l)}$ capsules via Eqn. (\ref{eqn:message_aggregate_u}) and Eqn. (\ref{eqn:message_aggregate_i}), respectively.
\ENDFOR
\STATE Merge dynamic routing:
\STATE Initialize coupling coefficients $b_{p}$.
\FOR{$\theta = 0 \rightarrow \theta$}
    \STATE Utilize dynamic routing to obtain the account capsule $\tilde{\mathbf{C}}_{A_k}^{(\theta)}$ via Eqn. (\ref{eqn:dynamic_routing}).
    \STATE Update coupling coefficients $b^{(\theta)}_p$ by Eqn. (\ref{eqn:update_ef}).
\ENDFOR
\STATE Obtain final representations $\mathbf{E}_A$, $\mathbf{E}_S$ by Eqn. (\ref{eqn:concat_cap}).
\end{algorithmic}
\end{algorithm}

Algorithm \ref{alg:sa} shows the pseudo-codes for the Subspace Alignment (SA) which is another key component in LightGC$^2$N.
\begin{algorithm}
\caption{Subspace Alignment}
\label{alg:sa}
\textbf{Input}: Input the sequence embeddings $\mathbf{E}_S$ obtained by GC$^2$N.

\textbf{Output}: Output the sequence embeddings $\hat{\mathbf{E}}_S$ refined by SA.
\begin{algorithmic}[1] %[1] enables line numbers
\STATE use K-means on $\mathbf{E}_S$ to initialize the subspace bases $\mathbf{D}=\{\mathbf{d}_1, \mathbf{d}_2, \dots, \mathbf{d}_j, \dots, \mathbf{d}_{\alpha}\}$.
\FOR{$S_k \in \mathcal{S}$}
    \STATE Calculate subspace affinities $s_{ij}$ between sequence embeddings $\mathbf{E}_{S_k}$ and subspace bases of $\mathbf{D}$ by Eqn. (\ref{eqn:sub_aff}).
    \STATE  Apply the subspace affinities $s_{ij}$ to refine the sequence embeddings $\mathbf{E}_{S_k}$ via Eqn. (\ref{eqn:refine_seq}).
    \STATE Calculate InfoNCE loss $\mathcal{L}_{C}^{S_k}$ between the refined representations $\mathbf{Z}_{S_k}$ with the normalized sequence embeddings $\mathbf{\hat{E}}_{S_k}$ by Eqn. (\ref{eqn:ssl_sub}).
\ENDFOR
\STATE Obtain the overall contrastive loss $\mathcal{L}_{C}$ by Eqn. (\ref{eqn:ssl_all}).
\STATE Output the refined sequence embeddings $\hat{E}_S$ by Eqn. (\ref{eqn:sum_sequence}).
\end{algorithmic}
\end{algorithm}

\subsection{A5. Theoretical Foundation Analysis}
\subsubsection{Graph Capsule Convolutional Network.} In GC$^2$N, graph capsule convolutional networks are present to capture the fine-grained associations between interactions and latent users within shared-account sequences. Unlike traditional graph neural networks that treat each interaction as an independent node, Capsule Graphs represent each interaction as a capsule, which encodes both the local and global information of the interaction \cite{Zheng_Capsule_Graph_22_SIGIR}. Capsule embeddings facilitate the representation of nodes with more complex internal structures and higher-dimensional outputs \cite{Wu_Capsule_Graph_23_TNNLs}. This advancement in embedding technology permits the model to effectively identify and cater to the nuanced preferences of users within a shared account, who may exhibit diverse interests. The dynamic routing mechanism inherent in graph capsule convolutional networks enables the message propagation between primary and senior capsules, contingent upon their semantic similarities. This functionality empowers the model to discern and learn the intricate relationships between various accounts and their respective latent users, thereby accurately ascertaining the individual contributions of each latent user to the overall account preference.
\subsubsection{Subspace Alignment.} Subspace Alignment (SA) is a crucial component of LightGC$^2$N that aims to refine sequence representations and align them with the preferences of latent users. SA utilizes low-rank subspace bases to cluster the hybrid preferences of multiple latent users within a shared account.
The low-rank representation hypothesis maintains that complex \cite{Xie_Sub_22_ICDE}, high-dimensional data is able to be represented by a limited number of lower-dimensional subspaces \cite{Zhang_Sub_2018_WISE}. In the context of shared-account sequential recommendation, this assumption holds true as the preferences of an account consist of multiple distinct latent users' preferences.
Traditional clustering methods based on the self-representation matrices operate in the original high-dimensional space and suffer from high computational complexity issues \cite{Cai_subspace_22_CVPR}. SA projects the sequence embeddings into the low-dimensional subspace defined by the bases, making the clustering process more efficient. 
The subspace bases provide interpretable insights into the preferences of different latent users. Each base represents a distinct subspace of a latent user, and the corresponding affinity values indicate the degree to which a sequential preference belongs to that latent user.

\subsection{A6. Additional Ablation Studies}
\begin{figure}[h]
\centering
\includegraphics[width=1.0\columnwidth]{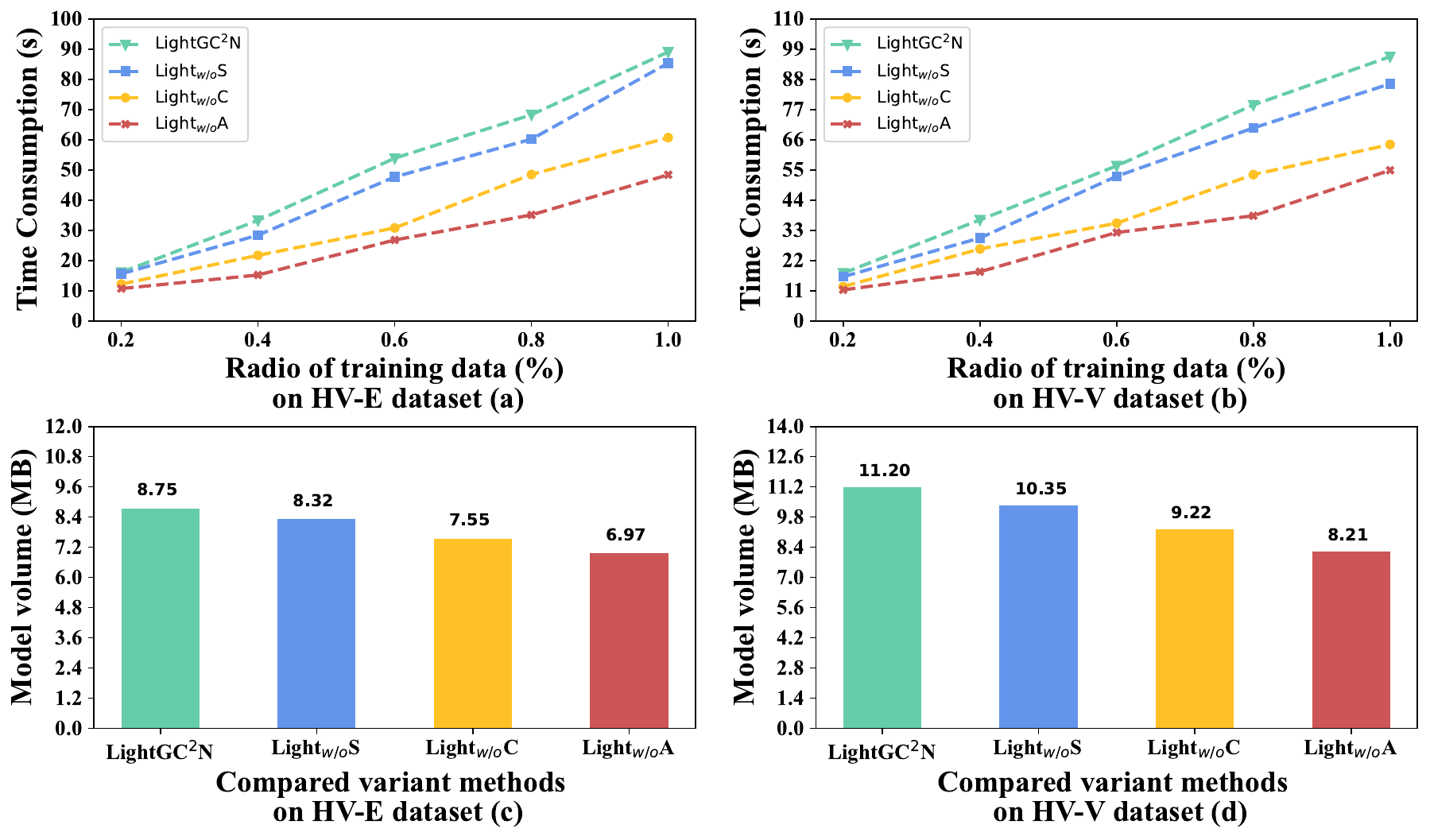} % Reduce the figure size so that it is slightly narrower than the column. Don't use precise values for figure width.This setup will avoid overfull boxes.
\caption{Comparison of time consumption and parameter scale between LightGC$^2$N and its variants.}
\label{fig5:efficiency_appendix}
\end{figure}

We conduct additional ablation studies on HV-E and HV-V to further investigate the contributions of the key components to the lightweight performance of LightGC$^2$N. Specifically, we further compare the training consumption and parameter scale between LightGC$^2$N and its three most significant variants (i.e., Light$_{w/o\text{C}}$, Light$_{w/o\text{S}}$, and Light$_{w/o\text{A}}$). The experimental results on the HV-E and HV-V datasets are presented in Figure \ref{fig5:efficiency_appendix}. The observations are summarized as follows.
1) As shown in Figure \ref{fig5:efficiency_appendix} (a) and (b), LightGC$^2$N exhibits more stable performance than the other variant methods under different ratios of training data, indicating that the core components of LightGC$^2$N are more scalable when dealing with large-scale datasets.
2) From Figure \ref{fig5:efficiency_appendix} (c) and (d), it can be observed that both removing the GC$^2$N (i.e., Light$_{w/o\text{C}}$) and the SA (i.e., Light$_{w/o\text{S}}$) have little impact on the model parameter scale. This observation further demonstrates the contribution of each key component to the lightweight performance of LightGC$^2$N.

\subsection{A7. Additional Hyper-parameters Analysis}
\begin{figure}[h]
    \centering
    \begin{minipage}[t]{1\linewidth}
        \centering
        \includegraphics[width=1.0\columnwidth]{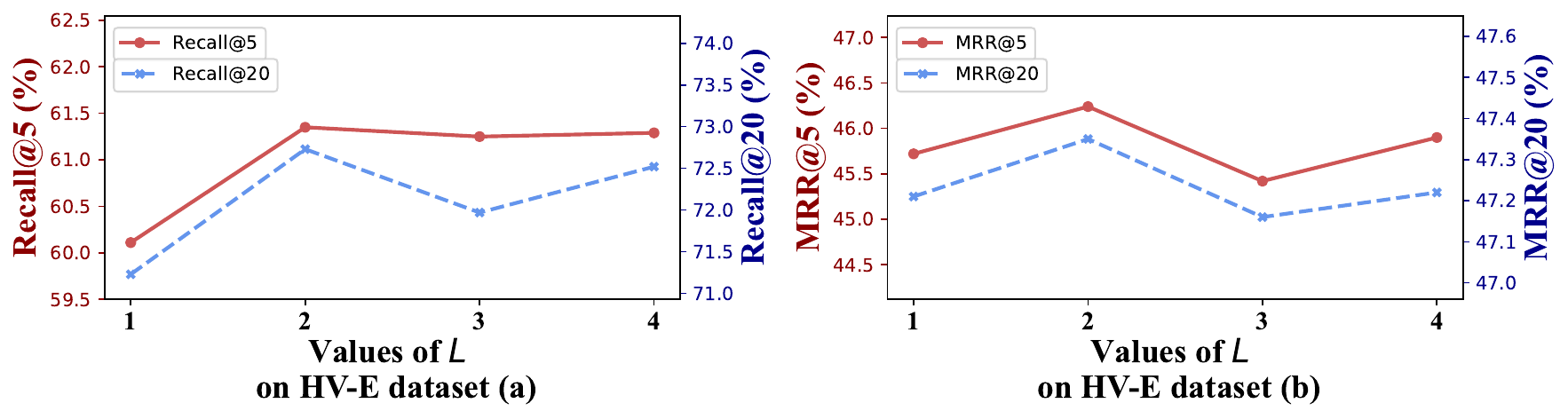}
    \end{minipage}
    \begin{minipage}[t]{1\linewidth}
        \centering
        \includegraphics[width=1.0\columnwidth]{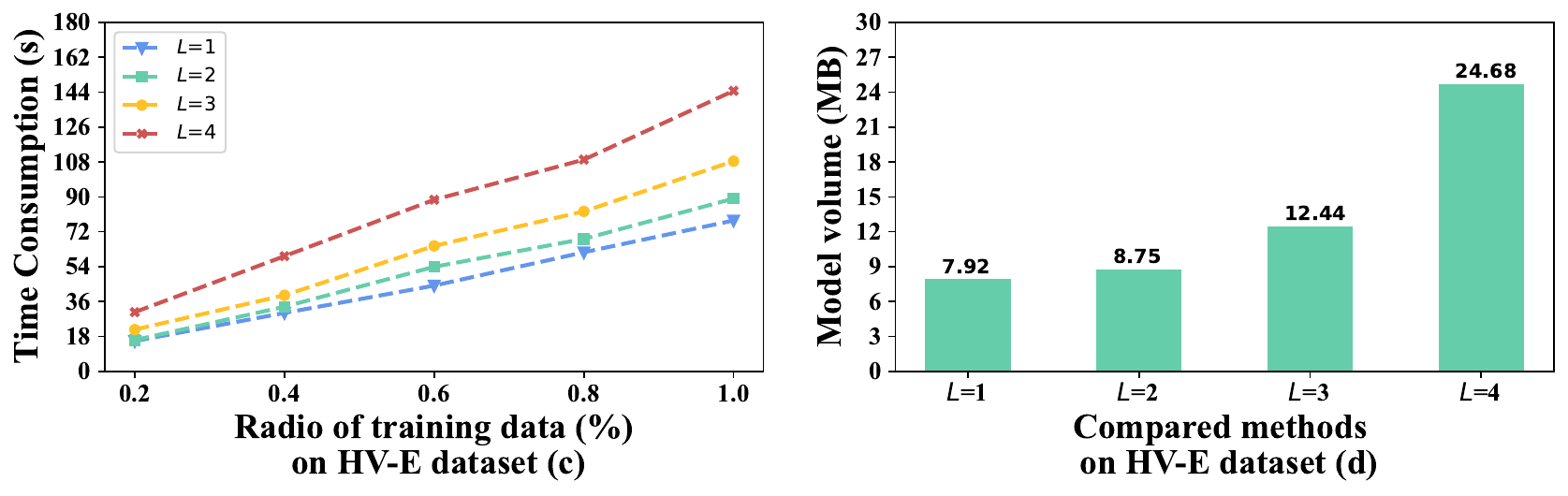}
    \end{minipage}
    \caption{Impact of hyper-parameter $L$ on the HV-E dataset.}
    \label{fig6:hyper_param_appendix}
\end{figure}

In this section, we conduct a series of experiments to investigate the impact of another significant hyper-parameter $L$, which controls the number of layers of the graph convolutions on the primary capsule graphs. The experimental results are shown in Figure \ref{fig6:hyper_param_appendix}.

As demonstrated in Figure \ref{fig6:hyper_param_appendix} (a) and (b), LightGC$^2$N achieves its best performance when $L$ is set to 2. After $L = 2$, an increase in $L$ is observed to be accompanied by a fluctuating decline in the model's performance. This observation indicates that the graph convolutional operations conducted on the primary capsule graphs are still subject to the over-smoothing issue. Additionally, as depicted in Figure \ref{fig6:hyper_param_appendix} (c) and (d), both the time consumption and the parameter scale of LightGC$^2$N exhibit continuous growth with the increasing number of convolutional layers $L$, and the rate of increase accelerates as the number of layers increases. Hence, we set $L$ to 2 in our experiments to maintain a balance between prediction accuracy and training consumption.

\end{document}